\documentclass[sigconf]{acmart}
\renewcommand\footnotetextcopyrightpermission[1]{}
\usepackage{booktabs} 
\usepackage{subfigure}

\setcopyright{rightsretained}

\copyrightyear{2017}
\acmYear{2017}
\setcopyright{rightsretained}
\acmConference{RecSysKTL Workshop @ ACM RecSys '17}{}{August 27, 2017, Como, Italy}
\acmDOI{N/A}
\acmISBN{}
\acmPrice{}
\begin{document}

\title{Multi-Stakeholder Recommendation: Applications and Challenges}

\author{Yong Zheng}
\orcid{1234-5678-9012}
\affiliation{%
  \institution{School of Applied Technology}
  \streetaddress{Illinois Institute of Technology}
  \city{Chicago}
  \state{Illinois}
  \postcode{60616}
}
\email{yzheng66@iit.edu}

\renewcommand{\shortauthors}{Y. Zheng}

\begin{abstract}
Recommender systems have been successfully applied to assist decision making by producing a list of item recommendations tailored to user preferences. Traditional recommender systems only focus on optimizing the utility of the end users who are the receiver of the recommendations. By contrast, multi-stakeholder recommendation attempts to generate recommendations that satisfy the needs of both the end users and other parties or stakeholders. This paper provides an overview and discussion about the multi-stakeholder recommendations from the perspective of practical applications, available data sets, corresponding research challenges and potential solutions.
\end{abstract}

%
%



\maketitle

\section{Introduction}
Recommender system is one of the information systems which assist user's decision making by recommending a list of appropriate items to the end users tailored to their preferences. It has been successfully applied to a number of applications, such as e-commerce (e.g., Amazon, eBay), online streaming (e.g., Netflix, Pandora), social networks (e.g., Facebook, Twitter), tourism (e.g., Tripadvisor) and restaurant (e.g., Yelp), etc.

{\let\thefootnote\relax\footnote{Presented at the 2017 Workshop on Value-Aware and Multistakeholder Recommendation.}}

Traditionally, recommender systems produce a list of recommendations to match the user preferences. For example, a learning-based recommendation algorithms may be developed to minimize the prediction errors or maximize the top-$N$ recommendations. These optimizations only takes the utility of the end users into account. However, the receiver of the recommendations may not be the only stakeholder in the system. For example, in the dating application, a young man may prefer a recommended dating woman who also wants to date with him, rather than only the pool of ladies that the young man likes. This is a well-known case in the reciprocal recommendation~\cite{pizzato2010reciprocal,pizzato2010recon,xia2015reciprocal}, where it is necessary to consider the utilities of two parties to produce accurate recommendations. In the advertising area, not only the user preferences but also the interests of the advertisers should be taken into account. For example, the advertiser would like to present the car advertisement to the appropriate customers, rather than any groups of the viewers who like cars. Young kids may like cars too but they may not have the capability to make purchases, which decreases the utility of the advertisers in this example.

The topic of multi-stakeholder recommender systems~\cite{burke2016towards} was raised recently. The idea behind is that the perspectives and utilities of multi-stakeholders are useful to be incorporated into the  recommendation process. The quality of the item recommendations should also depend on the utility of other stakeholders rather than the end user themselves.

Due to that fact that multi-stakeholder recommendation is still an early-stage research direction, this paper focuses more on the introduction and discussions about the potential applications, available data sets, corresponding research challenges and solutions.

\section{Related Work}
There are several research topics which are closely related to the optimization based on multiple utilities, though multi-stakeholder recommendation is viewed as a novel research topic. We introduce and discuss these related work in this section. Some of these work or techniques could be reused to solve the problems in the multi-stakeholder recommendations.

\subsection{Reciprocal Recommendation}
In a reciprocal recommender, the user and the item have similar standing in the system, in that both have preferences that must be satisfied~\cite{pizzato2010reciprocal}. More specifically, both the user and the item models represent people. Such "people-to-people" recommendations have been applied to several areas, including social networks~\cite{guy2015social}, online dating~\cite{pizzato2010recon,xia2015reciprocal}, employment and recruitment~\cite{yu2011reciprocal,mine2013reciprocal}.

For example, the social tie in Facebook can only be established by the approval of the two users. In the job market, a job offer can only be issued when both the employer and the employee are satisfied. However, there are only two parties involved in these applications, and each stakeholder may have the same standing or similar utilities to be optimized. For example, a successful recommendation in online dating should be the situation that the receiver of the recommendation prefers the suggested mate, while the recommended person may also like the receiver of the recommendations. Therefore, reciprocal recommendation can be considered as a special case in multi-stakeholder recommender system.

\subsection{Multi-Criteria and Group Recommendation}
Multi-criteria recommender systems~\cite{manouselis2007analysis} is another type of the recommender systems that produce recommendations by integrating multi-criteria preferences. Take TripAdvisor for example, a user may leave ratings on different aspects of the hotels, such as room size, cleanness, room service, business supports, in addition to the user's overall rating on the hotel. Each criterion actually can be viewed as one of the utilities from the perspective of the users. The challenge becomes how to take these multiple utilities into consideration to help a user select appropriate hotels. Apparently, there is only one stakeholder considered in the multi-criteria recommender systems, but the multiple criteria preferences can also be reused and extended to other stakeholders. The optimization approach in the multi-criteria recommendations may also be reused to serve the multi-stakeholder recommendations.

Group recommender systems~\cite{masthoff2011group}, by contrast, try to produce recommendations for a group of people. The challenge behind is that each individual may have their own tastes on the items, which leads to the difficulty of the preference representations at the group level. Therefore, the utilities at the group level and the individual levels should be considered in the recommendation process. The research on group recommendations is also correlated with the multi-stakeholder recommendations: on one hand, it is interesting to view each individual and each group as multi-stakeholders, and the group recommendation strategies, such as least misery, most pleasure, fairness, could also be useful in assisting multi-stakeholder recommendations. On the other hand, the level of groups may not only exist on the side of end users, but also happen on other stakeholders. For example, a group of tourists may consider to purchase a bundle of admission tickets to visit multiple museums. The museums on the bundle tickets can be viewed as another group, while there could be different perspectives of the museums. For example, the museums for educational purpose may prefer more students as their visitors. Or, the museums only accept students as visitors in specific time period.

\subsection{Multi-Objective Recommendation}
Utility optimization is one of the challenges in multi-stakeholder recommendations. Without considering multiple stakeholders, multi-objective recommendations~\cite{ribeiro2012pareto,rodriguez2012multiple} have been explored in the traditional recommender systems, such as the work that additionally consider novelty and diversity~\cite{vargas2011rank} in the recommender systems, or the recent work about fairness aware recommendations~\cite{modani2017fairness,serbos2017fairness}. The existing multi-objective optimization approaches could be reused to serve the multi-stakeholder recommendations, while the "objectives" are no longer limited to the perspectives of the end users.

\section{Utility Representations}~\label{sec:utilityrep}
In this section, we discuss different representations of the utilities that can be used as the optimization objectives in the multi-stakeholder recommender systems.

\subsection{Existing Preferences}
The simplest and most straightforward way to represent the utility of multiple stakeholders is to reuse the existing preferences in the system. Take E-commerce applications for example, the buyers can give ratings to the items they purchased, while the sellers may also leave ratings to the buyers, such as eBay.com. In addition to the numerical ratings, other types of preferences can also be utilized, such as the binary feedback based whether a purchase is made or not, the user's click-through data or browsing behaviors on the Web, etc. Existing preferences, whatever explicit or implicit ones, can be used to represent the utility of stakeholders. Burke, et al.~\cite{burke2016towards} made the attempt to reuse the existing preferences to calculate the gain values which are used to represent the utilities in the recommender systems.

\subsection{Utility Models}
In addition, it is also possible to build more complicated utility models to represent the utilities. The the movie watching for example, the frequency of movie genres that are associated with a user's most recent favorite videos can be used to infer the user's preferences or utilities, if we simply make the assumption that a user may like the same type of the movies they like recently. This is a simple time-dependent movie genre model, but there could be more complicated utility models. The idea behind is to aggregate the existing preferences to build the utility model, while the effectiveness of the utility model should be evaluated beforehand. For example, we may need to validate whether a user usually likes to watch similar type of the movies they like recently in the system.

\subsection{Other Representations}
As mentioned previously, a user's preferences on multiple aspects of the items, i.e., the multi-criteria preferences, can also be viewed as user utilities. The multi-criteria preferences could be collected from not only the end users, but also other stakeholders. Similarly, the preferences at the individual or group levels in the group recommender systems can also be utilized to represent utilities. Again, the group may be created for not only the receiver of the recommendations, but also other stakeholders.

\section{Applications and Data Sets}
In this section, we categorize the multi-stakeholder recommendation into different applications, and discuss corresponding available data sets to collect or use. Basically, we believe the following factors are important to be taken into account:

\subsection{Reciprocal or Pluralistic Stakeholders}
We limit the notion of "reciprocal" to the situation that there are two peer stakeholders and they must share the utility at the same standing. For example, user $u_1$ would like to add user $u_2$ as friends on Facebook.com. They are peer since all of them are the users on Facebook. The utility for them to add others as friends may be the same. Similar facts can be found in the application of online dating. By contrast, the application of pluralistic stakeholders refer to the situation that there are two or more than two stakeholders, and the stakeholders may have distinct utilities. The advertising case as mentioned previously is one of the examples, where there are two stakeholders, but they may have different utilities. More generally, there are many other applications which may involve more than two stakeholders. For example, in the e-commerce system, the stakeholders are not only the users who make purchases, but also the sellers, the producers and the shipping companies.

\subsubsection{Available Data Sets}
In case of the reciprocal recommendation, social network and online dating data sets are available to use, such as the speed dating data\footnote{https://www.kaggle.com/annavictoria/speed-dating-experiment}, and network data collections by Stanford\footnote{https://snap.stanford.edu/data/}. It could be difficult to collect the advertising data, especially when it comes to the utilities from the perspective of the advertisers. The same thing may happen to the case of pluralistic stakeholders.

Educational learning could be a good case study in multi-stakeholder recommender systems. We are collecting our own data based on student projects in the academia. Each \textit{instructors} may assign group projects to the students in each course. Given a list of potential topics for the projects, each student should read the description of the projects, and rate the project from different perspectives (i.e., multi-criteria ratings), such as how difficult the data processing is, how appropriate the topic is, how popular the application area is, and how challenging the project is. Meantime, the instructors will also give multi-criteria ratings to each project, such as the level of difficulties, the degree of fitness for the students, the relations to industry applications or experiences, etc. These multi-criteria ratings can be viewed as the utilities for instructors and students respectively. In addition, teaching assistants may be another type of the stakeholders in this example, since they may grade students' work and their utilities are also relevant to the system. For example, the grading procedure of a programming project may be different from the one for a data analytics project. Currently, this project is undergoing and the data is relatively small right now. We expect a larger data that could be available for research in the next year.

\subsection{Single or Multiple Utilities}
There could be a single utility or multiple utilities for each stakeholder. The examples of social networks and online dating may be the ones where there is a single utility for each type of stakeholder. However, there could be multiple utilities rather than a single one for each stakeholder. The multi-criteria recommender system is a good example, where the preference on each criterion can be viewed as an individual utility. More generally, multiple utilities can be produced by the utility representations that are discussed in Section~\ref{sec:utilityrep}.

\subsubsection{Available Data Sets}
Accordingly, any data sets with multi-criteria preferences can be reused in the multi-stakeholder recommendations, such as the TripAdvisor data and Yahoo!Movie data used in~\cite{jannach2014leveraging}. However, we may only have the utilities of the end users, but miss the ones by other stakeholders. We can produce utilities for other stakeholders by utilizing existing preferences. For example, Burke, et al.~\cite{burke2016towards} utilized the MovieLens data and simulate the utilities for other stakeholders such as movie suppliers based on the user's ratings. Similar operations can be applied to the multi-criteria rating data for the purpose of the multi-stakeholder recommendations.

\subsection{Onefold or Correlated Stakeholders}
Sometimes, the relationships among stakeholders are not that as simple as the ones like user-to-user on social networks or buyer-to-seller in the e-commerce applications. But there may be complicated relations among them. For example, group recommendation is one of the examples, stakeholders may be at the group level (i.e., a group of the users) or individual level.
Furthermore, individuals may be associated with other relationships, such as whether they are connected over the Facebook, teacher-student or employee-supervisor relationships, which may result in a hierarchical structure or connections over networks. These relations may potentially influence the utilities from individuals to individuals. For example, due to the limited budget, the employee has to purchase the same type of the equipment as the ones bought by their supervisor. The utility from the perspective of employee yields to the supervisor's choices. Another example could be the information sharing at the social networks. The information that a student shares with his or her friends may be different from the ones that are shared with his or her instructors. Students may not share complaints about the teaching with their instructors. Therefore, it is necessary to take these relationships into consideration in the recommendation process.

\subsubsection{Available Data Sets}
The data sets used in the group recommendations can be reused in this case. For example, Masthoff~\cite{masthoff2003modeling} simulated a data based on the MovieLens movie data set. Liu~\cite{liu2012exploring} et al. used a data collected from Meetup.com. Our data which collects multi-criteria ratings on the student projects could be another option, since students may work in a team, and the group of the students in a team needs to leave ratings accordingly.

In terms of the correlations among the stakeholders, it may be useful to take user connections, such as social ties, into consideration. We may have to mine the relationships if they are not explicitly known in the applications. For example, text mining on Tweets could be used to infer the instructor-student relationships on Twitter.

\section{Challenges and Solutions}
Based on the different applications of the multi-stakeholder recommender systems discuss in Section 4, we can generate a list of research challenges as follows.

\subsection{Utility Optimizations}
The most straightforward research direction is that how to optimize the recommendations by taking multi-stakeholder's utilities into considerations. The challenge behind is that the increment on one utility may hurt other utilities. To simplify the problem, let's take the following situation for example in our discussions: there are two types of the stakeholders and there is a single utility for each stakeholder.

Evolutionary algorithms are one of the most popular techniques in multi-objective optimizations~\cite{deb1999multi,ribeiro2012pareto}. The most straightforward idea is to update the factors in the recommendation algorithms when we did not hurt one of the utilities significantly. For example, optimizing utility $t_2$ may hurt the value in utility $t_1$. We can build a recommendation algorithm to optimize a single utility $t_1$. Then we set a tolerance to make sure the $t_1$ can be decreased but up to a percentage $a$\%. Evolutionary algorithm, afterwards, can be integrated into the recommendation models (such as matrix factorization) to update the factors and optimize $t_2$. The latent factors will only be updated when it improves $t_2$ and does not downgrade $t_1$ more than $a$\%. These approaches could be very useful, especially when there are conflicting utilities.

One of other potential solutions is to come up new metrics which unify and integrate multiple utilities. Vargas, et al.~\cite{vargas2011rank} proposed a framework for the definition of novelty and diversity metrics in the recommender systems that unifies and generalizes several state-of-the-art metrics. F-measure in information retrieval can be considered as another example of combining precision and recall metrics. Similar operations can be applied to the utility functions in the multi-stakeholder recommendations. The only difference is that we need to consider the utilities from multiple stakeholders, rather than only the utilities from the receiver of the recommendations. However, the optimization is more challenging if there are conflicting interests among the multi-stakeholders. Rodriguez, et al.~\cite{rodriguez2012multiple} proposed an optimizing framework to add competing objectives one by one and the framework allows for fine control over any potential loss in relevance as additional aspects of the system's overall utility function are optimized by using job or employment as case studies.

Recently, the notion of "fairness" has been introduced to the classification~\cite{zemel2013learning} and group recommender systems~\cite{serbos2017fairness} in order to obtain both group fairness and individual fairness. Group fairness in the classification problem, ensures that the overall proportion members in a protected group receiving positive classification are identical to the proportion of the population as a whole~\cite{zemel2013learning}. Serbos, et al.~\cite{serbos2017fairness} redefines the fairness to produce a package of items to a group of the users in order to make sure at least one item satisfies each user in the group. Abdollahpouri,et al.~\cite{himan2017recsys} developed a recommendation model to leverage the popularity bias in the learning-to-rank recommendations by taking advantage of the predefined intra-list binary unfairness. The research on fairness-aware recommendations is closely related to multi-objective optimizations and it is promising to be utilized as the utility optimization solution in the multi-stakeholder recommender systems as well.

\subsection{Correlations Among Utilities}
The correlation among utilities may be necessary to be considered in the recommendation model. The assumption behind is that the value in one utility may affect the value in other utilities.

\begin{figure}[ht!]
\centering
\includegraphics[scale=0.85]{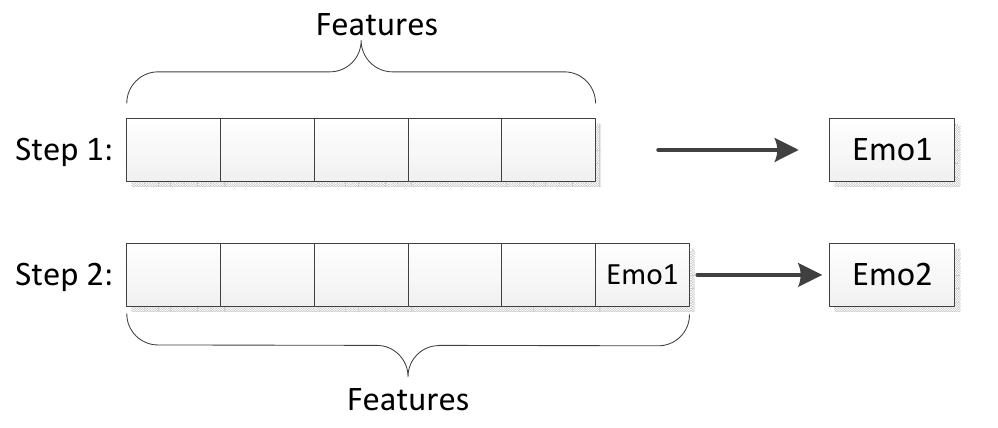}
\caption{Workflow of the Chain}
\label{fig:chain}
\end{figure}

Take the multi-criteria recommendation in the case of hotel booking on TripAdvisor.com for example, the room size may be a key factor for a user's decision on hotel bookings in terms of a family trip. A low rating on "room size" may directly affect the user's rating on another criterion ``value", as well as the user's overall rating on the hotel. To incorporate these correlations into the recommendation model, Sahoo, et al.~\cite{sahoo2012research}, builds probabilistic recommendation algorithms based on the pre-defined graphical relationships. We propose one approach, "CriteriaChains"~\cite{zheng2017criteria} which predicts the utility values one by one in the shape of chains. We further apply the model to predict a user's emotional states~\cite{zheng2017emo}. The approach can be simply described by Figure~\ref{fig:chain}. Assume we have two emotional variables to predict. First of all, we utilize the pre-defined features (such as user demographic information and/or item features) to predict the value in the first emotional variable. Then the predicted value will be viewed as additional feature to make predictions for the second emotional state. By making predictions in a chain, the correlations among these emotional variables are taken into account.

These models could be extended to explore the correlations among different utilities in multi-stakeholder recommendations. Not only the multiple utilities from one stakeholder, but the ones from different stakeholders can be taken into consideration.

Furthermore, the correlations among the utilities could be more complicated. Take the case of student project for example, students may want to choose easy ones as their final project, while instructors hope they can select more challenging ones in order to examine student's skills and give them a higher grade. In this case, the correlations between these utilities, such as the ease of the project from a student's prospective, and how challenging the project is from the view of the instructor, turn out to be conflicting interests. Related techniques, such as game theory, may be useful to balance the utilities of different stakeholders.

\subsection{Relationships Among Multi-Stakeholders}
The user relationships may exist in the same type of the stakeholders, such as the group recommendations where the users in the same group may be related. In addition, the relationship may take effect among different stakeholders. Assume there are three stakeholders, \textit{instructor, grader} and \textit{students}, the grader may give a higher grader if he or she has a good relationship with a student. Another example could be the information sharing at the social networks. The information that a student shares with his or her friends may be different from the ones that are shared with his or her instructors. Students may not share complaints about the teaching with their instructors.

As far as we know, there are no existing work on this type of the research directions, even in the area of social recommendations. There are some work related to the effect by the social capitals~\cite{tsai1998social,kim2016effect} in the social networks, but the situation could be much more complicated in the multi-stakeholder recommendations. In our view, the relationships may produce some constraints or biases that are necessary to be utilized in order to build more effective recommendation models.

\subsection{Impact by Dynamic Factors}
The stakeholder's expectations may not be always the same, while they may change by some dynamic factors, such as the context factors or emotional states. Take the hotel recommendation for example, the room size may be a key factor if the user is travelling with family members. It may be not essential if the user travels alone. The same thing may happen by considering temporal effectors. Cartoon may be a user's favorite video when he or she was young. But the user prefers other types of the movies when he or she grew up. To build effective multi-stakeholder recommender systems, it is necessary to incorporate these dynamic factors (such as context) into the recommendation models. On one hand, the work in the context-aware recommendation models~\cite{zheng2014cslim,zheng2015simcars} may be helpful to improve the quality of the recommendations. On the other hand, the quality of multi-stakeholder recommendations should be evaluated based on different contextual situations.

\section{Conclusions and Future Work}
To some extent, some research problems in the multi-stakeholder recommendations are closely related to existing work, such as multi-objective and reciprocal recommendations. However, the multi-stakeholder recommendation could be more complicated when there are more stakeholders while there could be several utilities associated with each stakeholder, not to mention that there may be correlations among these utilities, as well as the stakeholder themselves. In this paper, we provide an overview of the different applications of multi-stakeholder recommendations. We further introduce available data sets and discuss potential research challenges and corresponding solutions. We expect open questions and welcome collaborations on the topic of multi-stakeholder recommendations.

\bibliographystyle{ACM-Reference-Format}
\bibliography{sigproc}

\end{document}